\documentclass[aps,prl,twocolumn,showpacs,superscriptaddress,amsmath,amssymb]{revtex4}
\usepackage{graphicx}
\usepackage{color}

\definecolor{Red}{rgb}{1.0,0.0,0.0}

\begin{document}

\title{Explosive metallic breakdown}

\author{Cl\'audio L. N. Oliveira}%
\email{lucas@fisica.ufc.br}%
\affiliation{Departamento de F\'isica, Universidade Federal do
  Cear\'a, 60451-970 Fortaleza, Cear\'a, Brazil}%

\author{Nuno A. M. Ara\'ujo}%
\email{nmaraujo@fc.ul.pt}%
\affiliation{Computational Physics, IfB, ETH Z\"urich, H\"onggerberg,
  CH-8093 Z\"urich, Switzerland}%
\affiliation{Departamento de F\'{\i}sica, Faculdade de Ci\^{e}ncias,
  Universidade de Lisboa, P-1749-016 Lisboa, Portugal, and Centro de
  F\'isica Te\'orica e Computacional, Universidade de Lisboa, Avenida
  Professor Gama Pinto 2, P-1649-003 Lisboa, Portugal}%

\author{Jos\'e S. Andrade Jr.}%
\email{soares@fisica.ufc.br}%
\affiliation{Departamento de F\'isica, Universidade Federal do
  Cear\'a, 60451-970 Fortaleza, Cear\'a, Brazil}%
\affiliation{Computational Physics, IfB, ETH Z\"urich, H\"onggerberg,
  CH-8093 Z\"urich, Switzerland}%

\author{Hans J. Herrmann}%
\email{hans@ifb.baug.ethz.ch}%
\affiliation{Computational Physics, IfB, ETH Z\"urich, H\"onggerberg,
  CH-8093 Z\"urich, Switzerland}%
\affiliation{Departamento de F\'isica, Universidade Federal do
  Cear\'a, 60451-970 Fortaleza, Cear\'a, Brazil}%


\begin{abstract}
  We investigate the metallic breakdown of a substrate on which highly
  conducting particles are adsorbed and desorbed with a probability
  that depends on the local electric field. We find that, by tuning
  the relative strength $q$ of this dependence, the breakdown can
  change from continuous to explosive. Precisely, in the limit in
  which the adsorption probability is the same for any finite voltage
  drop, we can map our model exactly onto the $q$-state Potts model
  and thus the transition to a jump occurs at $q=4$. In another limit,
  where the adsorption probability becomes independent of the local
  field strength, the traditional bond percolation model is recovered.
  Our model is thus an example of a possible experimental realization
  exhibiting a truly discontinuous percolation transition.
\end{abstract}

\maketitle
  
One of the main problems in the manufacturing of Integrated Circuits
(IC), where millions of nanometric metallic and semiconductor devices
are placed on a substrate, is pollution with metallic dust, since it
can induce an electric breakdown, leading to malfunctioning and a
shorter life-time of the IC~\cite{Alam2002,Niemeyer1984,Verweij1996}.
The following question then arises, under which conditions and how
fast such a system does collapse. To answer this question, one needs
to take into account that the deposition of metallic particles is
hindered by the local electric field. This strongly non-linear
interplay between adsorption and the local geometry gives rise to
interesting phenomena that we will explore here with the help of a
rather simple model, which however can still capture the essential
physics of the problem. In particular, we find that the collapse can
either be continuous or explosive depending on the physical
parameters.

Here, we model the substrate as a $L\times L$ tilted square lattice
with periodic boundary conditions in one direction and a voltage drop
$V_0$ applied in the other direction. For simplicity, the bonds of the
lattice can be either associated with highly resistive elements or
metallic particles, with resistances $R=1$ and $0$, respectively.
Nodes connected by metallic bonds constitute metallic clusters and
thus have the same electric potential. Resistive bonds that connect
sites of the same metallic cluster are called {\it internal bonds},
while all other resistive bonds are called {\it merging bonds}.
Internal bonds do not feel any field and therefore metal dust can be
adsorbed on them with a probability $q$ times higher than on merging
bonds, where the factor $q>1$ describes the relative deposition
disadvantage due to the presence of the field. Additionally the
probability of adsorbing a metallic particle decreases monotonically
with the strength of the local field gradient $\Delta V$, namely, we
will assume here generically that the adsorption probability decays as
$1-(\Delta V/V_0)^\gamma$, where $\gamma$ is another adjustable
parameter. In the framework of this model, it follows that the
probability for metallic bonds to replace resistive bonds
(adsorption process) is given by,
\begin{equation}
  W=\dfrac{p}{q}\left[1 + (q-1)\left(1 - \left(\dfrac{\Delta V}{V_0}\right)^\gamma\right)\right].
\end{equation}
One should note that, if the bond is internal, the adsorption
probability becomes $W=p$, since $\Delta V=0$.
\begin{figure}[b]
\begin{center}
  \includegraphics*[width=\columnwidth]{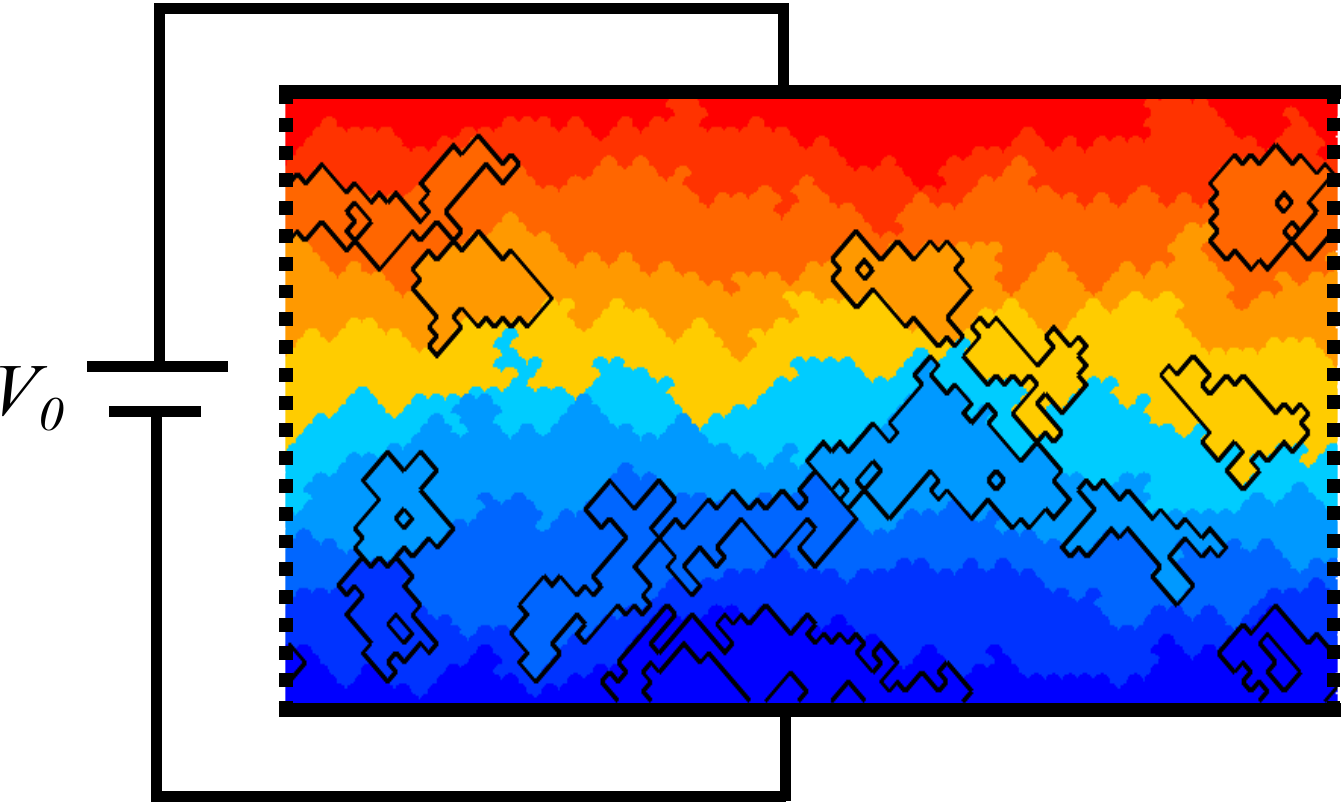}
\end{center}
\caption{(Color online) Plot showing a typical state of the system for
  $q=10$, $L=128$, $\gamma=0.1$, and $p=0.57$. A potential drop $V_0$
  is applied from top to bottom and periodic boundary conditions from
  left to right.  The nodes are colored according to their
  corresponding potential, with the constraint that nodes belonging to
  the same metallic cluster hold the same potential. Nodes with high
  potential are shown in red (top), while low potential ones are
  presented in blue (bottom). The boundaries of several metallic
  clusters are highlighted for better visualization.} \label{f.medium}
\end{figure}
\begin{figure*}[t]
  \begin{center}
    \includegraphics*[width=\textwidth]{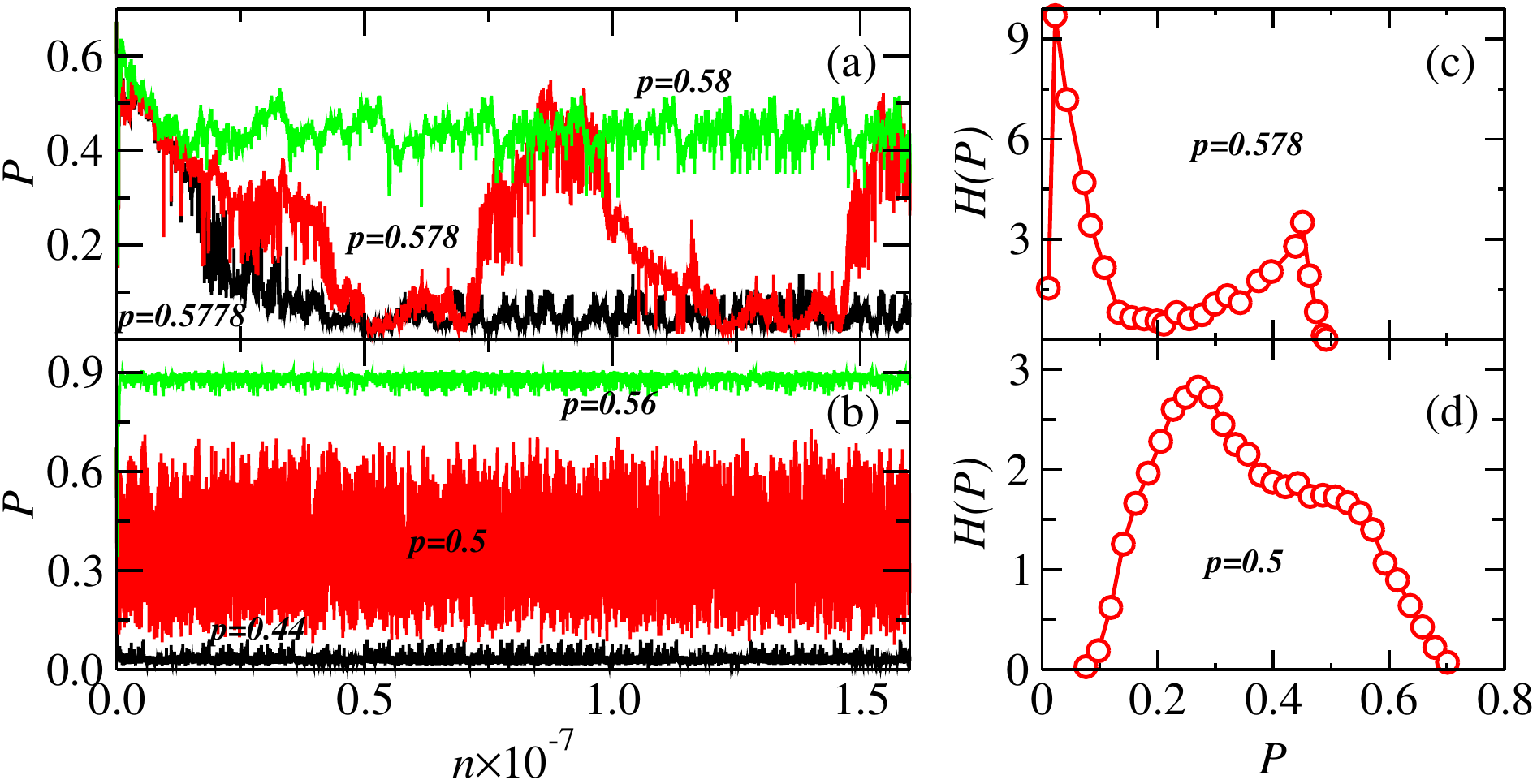}
  \end{center}
  \caption{(Color online) (a) Fraction $P$ of nodes in the largest
    cluster with the number of iterations $n$ (sampled bonds) for
    $q=10$, $L=128$, $\gamma=0.1$, and distinct values of $p$ below
    (black), above (green), and at the transition point $p_c=0.578$
    (red). (b) The same as in (a) but for $\gamma=1$ and $p_c=0.5$.
    The histograms for $P$ at the transition point are shown in (c)
    and (d) for $\gamma=0.1$ and $\gamma=1$, respectively. For
    $\gamma=0.1$, the histogram is bimodal and the evolution is
    characterized by metastability, two typical signs of a
    discontinuous transition. By contrast, for $\gamma=1$ the
    transition is continuous since the histogram is unimodal and there
    is no evidence of metastability.}
  \label{f.evolution}
\end{figure*}

The inverse process, namely desorption, then happens naturally with a
probability one minus the probability of adsorption: $1-p$. The model
has three parameters: $p$ is a measure for the amount of metallic
dust, $q$ is the enhancement of adsorption if there is no local
electric field, and $\gamma$ is the dependence on the strength of the
local field. For $\gamma=\infty$, all resistive bonds (merging and
internal) can be replaced with the same probability $p$ so that, in
this limit, one recovers classical bond
percolation~\cite{Stauffer1992}. Moreover, as we will discuss later,
one recovers the $q$-state Potts model for
$\gamma=0$~\cite{Potts1952,Gliozzi2002}. In a way, our model is the
inverse of a fuse
model~\cite{Kahng1988,Andrade2011,Pose2012,Moreira2012}.

The simulations are performed here by randomly choosing, at each
iteration, a bond between neighboring nodes $i$ and $j$ and attempting
to change its state according to the probabilities previously
described. Each time a merging bond is identified, the local potential
drop $|\Delta V|_{ij}$ is calculated by solving the Kirchhoff
equations for each node simultaneously. This is equivalent to solve
the Laplace equation $\nabla^2V=0$~\cite{HSL} by discretization, and
impose that nodes belonging to the same metallic cluster have the same
potential, as illustrated in Fig.~\ref{f.medium}.

\begin{figure}[t]
\begin{center}
  \includegraphics*[width=\columnwidth]{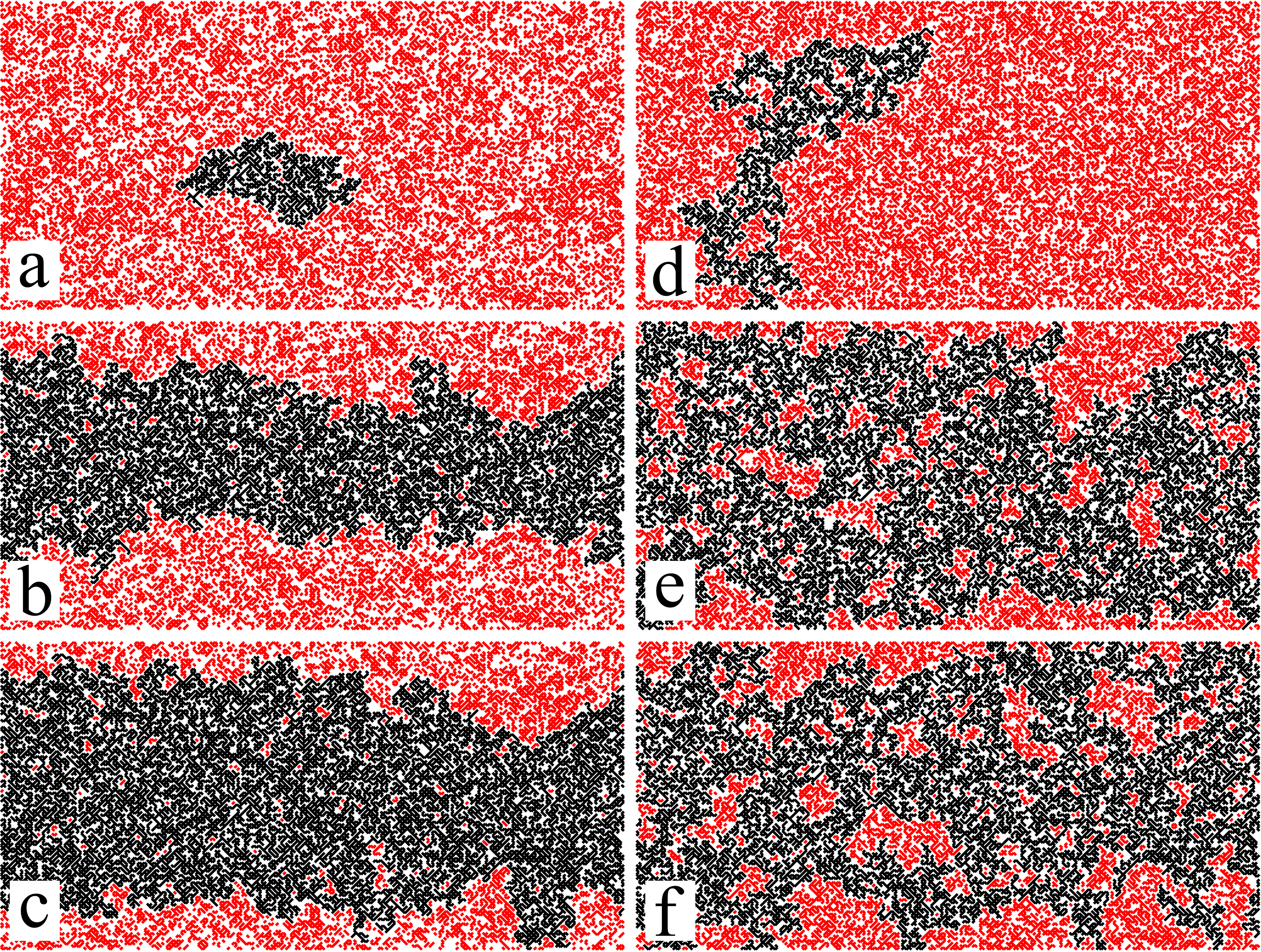}
\end{center}
\caption{(Color online) Snapshots of steady state configurations for
  $q=10$, $L=128$, $\gamma=0.1$, and (a) $p=0.57$, (b) $p=0.58$, and
  (c) $p=0.59$, and for $\gamma=1$ and (d) $p=0.49$, (e) $p=0.5$, and
  (f) $p=0.51$. Metallic bonds belonging to the largest cluster are
  colored in black, while the bonds belonging to other metallic
  cluster appear in red. For $\gamma=0.1$, the largest metallic
  cluster is compact and grows abruptly at the threshold, $p_c\approx
  0.58$, while for $\gamma=1$ it is fractal and grows rather smoothly
  with $p$.}
\label{f.snapshot}
\end{figure}

Regardless of the starting configuration, a steady state is reached
after a certain number of iterations. In Fig.~\ref{f.evolution} we
show how the fraction $P$ of nodes in the largest metallic cluster
fluctuates with the number of iterations $n$, at the steady state, for
$q=10$, and different values of $p$ and $\gamma$. For fixed values of
$q$ and $\gamma$, by increasing $p$ one reaches a value
$p_c(q,\gamma)$ at which the system electrically breaks down.  At this
point, a spanning metallic cluster can appear after steady state,
which makes the system fluctuate strongly between resistive and
metallic configurations (see Fig.~\ref{f.evolution}). In particular,
$p_c(q,\gamma=\infty)=1/2$ because, as previously stated, for
$\gamma=\infty$ our model recovers bond percolation on the square
lattice. For $\gamma=0.1$, as shown in Fig.~\ref{f.evolution}(a), $P$
mainly oscillates around two well-defined values, $P\approx 0.05$ and
$0.45$. The fact that the distribution of $P$ is bimodal (see
Fig.~\ref{f.evolution}(c)) indicates a metastability, namely, a clear
signature of a discontinuous transition. By contrast, for $\gamma=1$
(see Fig.~\ref{f.evolution}(b)), although the variable $P$ also
fluctuates around an average value, the distribution of $P$ is
unimodal and there is no sign of metastability (see
Fig.~\ref{f.evolution}(d)). Thus, the transition in this case is
continuous. Shown in Fig.~\ref{f.snapshot} are snapshots of steady
state configurations for both cases at and around the threshold $p_c$.
While for $\gamma=0.1$ a compact gigantic metallic cluster abruptly
appears leading to a discontinuous transition (see panels (a)-(c)),
for $\gamma=1$ the largest cluster looks fractal and grows with $p$
yielding a continuous transition (see panels (d)-(e)). As a
consequence, the metallic breakdown is either smooth or explosive
depending on the values of $\gamma$ and $q$.

\begin{figure}[t]
\begin{center}
  \includegraphics*[width=\columnwidth]{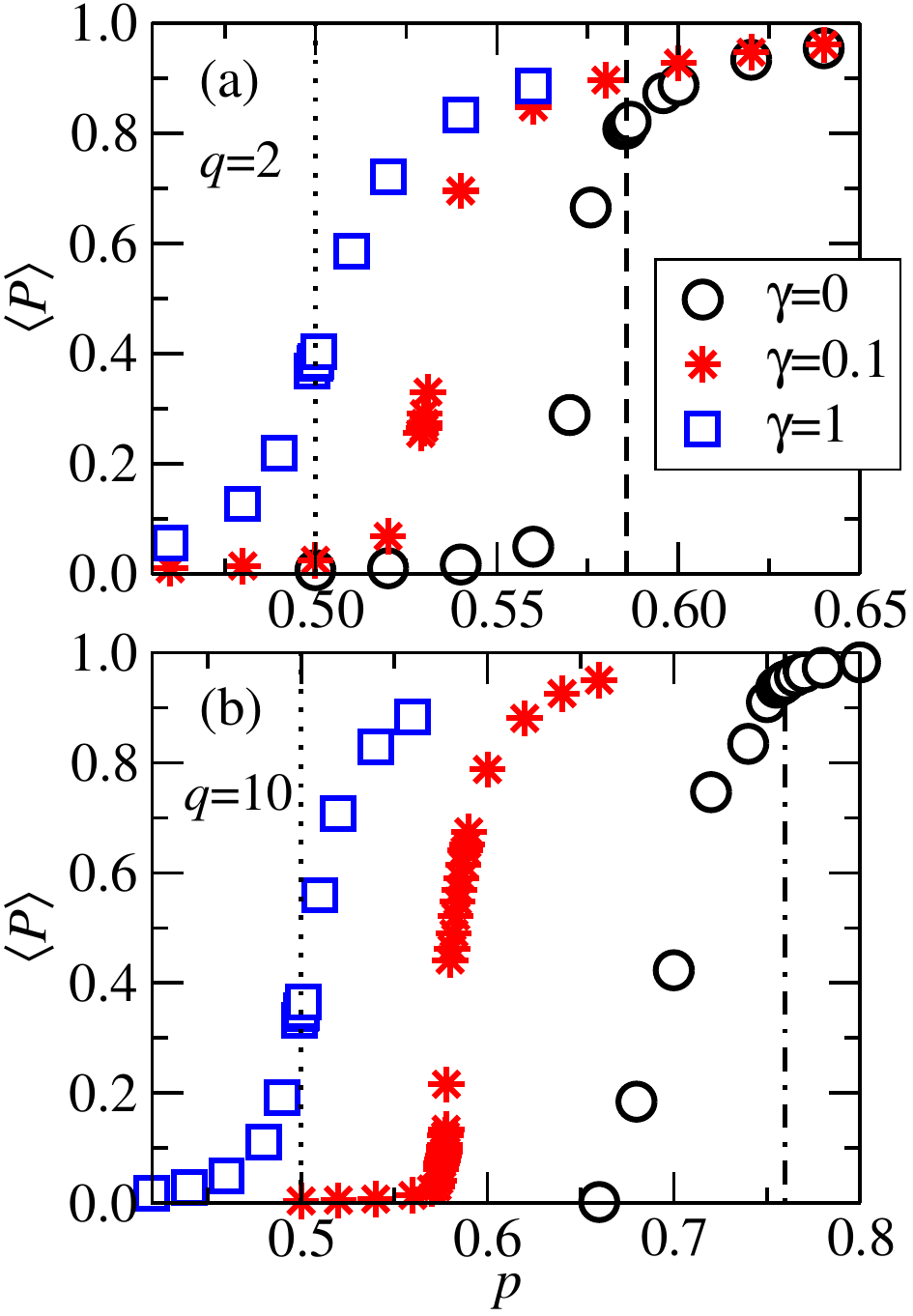}
\end{center}
\caption{(Color online) Dependence on $p$ the average fraction $\langle
  P\rangle$ of nodes in the largest cluster for $\gamma=0$,
  $0.1$, and $1$, for $q=10$ (a) and $q=2$ (b), with $L=128$. The
  transition point for a given $\gamma$ and $q$ is bounded between the
  critical point for bond percolation, $p_c(q=1,\gamma=0)=1/2$ (dotted
  lines), and the critical point for the $q$-state Potts model, given
  by Eq.~(\ref{e.pc}), where $p_c(q=2,\gamma=0)\approx0.5858$ (dashed
  line), and $p_c(q=10,\gamma=0)\approx0.7597$ (dotted-dashed line).
  \label{f.fracao}}
\end{figure}

The phase transition between resistive and metallic states is
described in terms of the average fraction of nodes in the largest
metallic cluster, $\langle P\rangle$, taken here as the order
parameter. For a given pair of $p$ and $\gamma$ values, we calculate
$\langle P\rangle$ by averaging $P$ over many iterations $n$ at the
steady state. Notice that the steady state may be reached with more or
less iterative steps depending on the parameters of the system. In
Fig.~\ref{f.evolution}(a), for example, not less than $3 \times
10^{6}$ steps were needed. In addition, well-defined histograms with
reduced fluctuations can only be obtained in some cases for $n>10^{7}$
steps. Since most of these steps involves the inversion of large
matrices, the calculation becomes prohibitively heavy for system sizes
$L>128$. In Fig.~\ref{f.fracao} we show $\langle P\rangle$ as a
function of $p$ for different values of $\gamma$, for $q=2$ in (a) and
$q=10$ in (b).  As depicted, the value of $p_c$ decreases with
$\gamma$ and increases with $q$, being always between the critical
points of bond percolation, $p_c=1/2$, and the one of the
corresponding $q$-state Potts model (vertical lines in the plots).
Moreover, these results also show that the order parameter $\langle
P\rangle$ becomes clearly steeper by increasing $q$.

In the particular case of $\gamma=0$, merging bonds become metallic
with probability $p/q$ and internal bonds with probability $p$. This
is precisely the Monte Carlo procedure to obtain the Coniglio-Klein
clusters~\cite{Coniglio1980} for the $q$-state Potts model as derived
by Gliozzi~\cite{Gliozzi2002,Wang2002} from the Kasteleyn-Fortuin
formulation~\cite{Fortuin1972,Wu1982}. Thus $\langle P\rangle$ is
nothing but the order parameter of the $q$-state Potts model for
$\gamma=0$. From self-duality, the transition point is known exactly
for the square lattice to be~\cite{Wu1982},
\begin{equation}\label{e.pc} 
p_c(q,\gamma=0)=\frac{\sqrt q}{1+\sqrt q}.
\end{equation}
Furthermore it is known that, in two dimensions, the phase transition
of the Potts model changes at $q=4$ from second order, for $q<4$, to
first order, for $q>4$. Our physical model then encompasses a plethora
of transitions between continuous and abrupt metallic breakdowns.
Remarkably, for the case $\gamma=0$, this is not a numerical but a
rigorous result. It constitutes in fact a beautiful example for
explosive (discontinuous) percolation~\cite{Baxter1973,Nienhuis1979}
which has an analytical approach, on one hand, and an experimental
realization, on the other.

\begin{figure}[t]
\begin{center}
  \includegraphics*[width=\columnwidth]{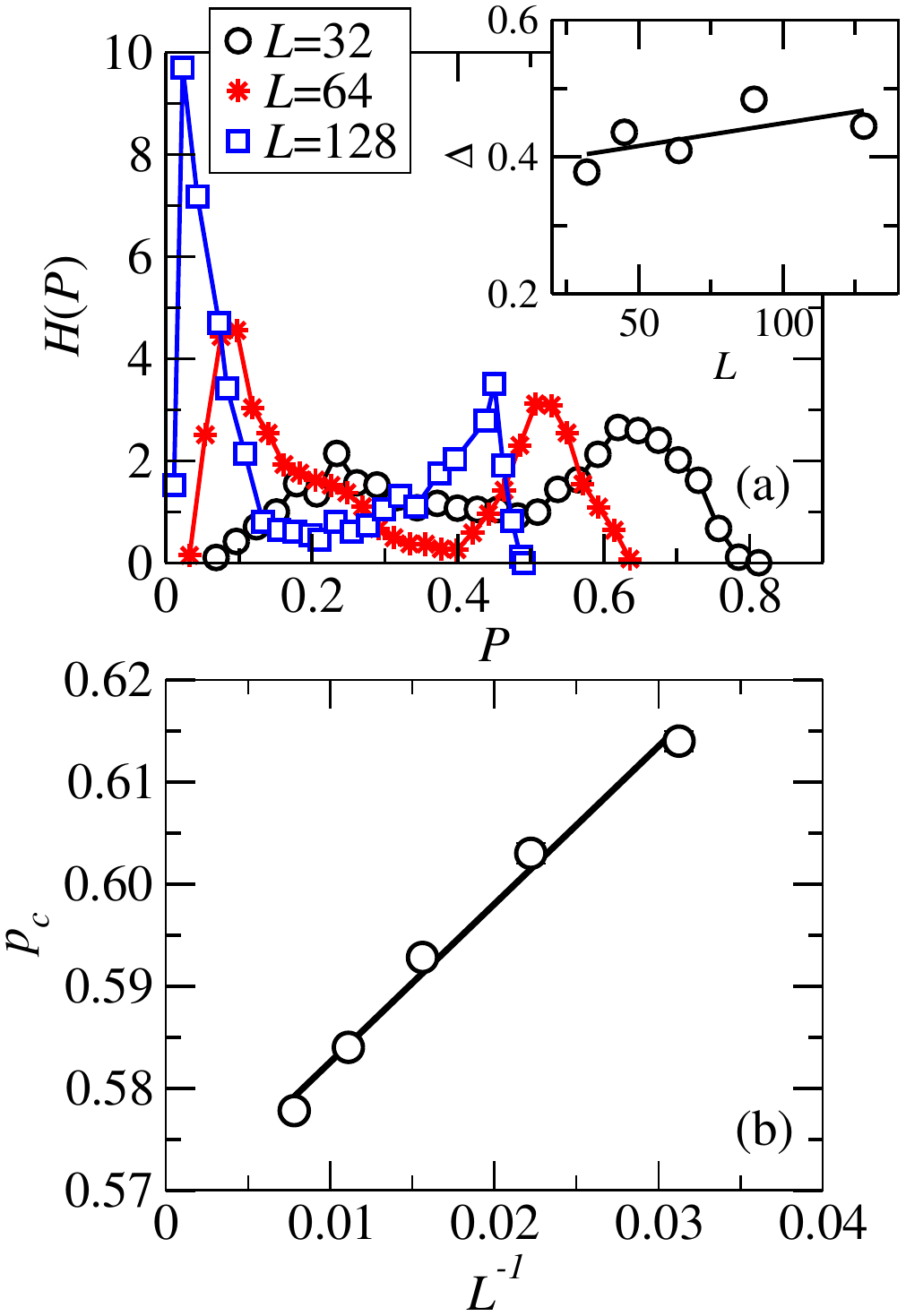}
\end{center}
\caption{(Color online) (a) Histogram of the fraction $P$ of nodes in
  the largest metallic cluster for different system sizes, $L$, at
  $p_c$ for $q=10$ and $\gamma=0.1$. In the inset of (a), $\Delta$ is
  the difference between the two peaks in the histogram. One clearly
  sees that $\Delta$ does not decrease with the system size, as one
  would expect for a continuous transition. For every $L$, $p_c(L)$ is
  obtained as the value of $p$ that maximizes the derivative of $P$,
  $\text{d}P/\text{d}p$. As shown in (b), $p_c$ can then be
  extrapolated to the thermodynamic limit through
  $p_c=0.5671+1.5463L^{-1}$.}
\label{f.hist}
\end{figure}

To see up to which point our result is also of more general
experimental validity, we investigate numerically if this transition
from continuous to abrupt exists for finite values of $\gamma$. The
two different patterns for the evolution of $P$ shown in
Fig.~\ref{f.evolution} provide strong evidence for this behavior,
although a more systematic study based on finite-size scaling is
necessary. One should note that, as shown in Fig.~\ref{f.fracao}, due
to the finite size of the simulated systems, the numerical data for
$\gamma=0$ (circles) still deviate considerably from those expected in
the thermodynamic limit, as given by Eq.~(\ref{e.pc}) (marked by
dashed lines). In particular, one can not recognize the predicted
continuous (discontinuous) transition in Fig.~\ref{f.fracao}(a)
(Fig.~\ref{f.fracao}(b)). Therefore we studied the histogram of the
order parameter at $p_c$ for different system sizes, as shown in
Fig.~\ref{f.hist} for the case $\gamma=0.1$ and $q=10$. Here $p_c$ was
determined from a careful finite-size extrapolation (see
Fig.~\ref{f.hist}(b)).  As clearly shown in Fig.~\ref{f.hist}, the
presence of a typical bimodal distribution and the fact that the
distance between peaks does not vanish in the thermodynamic limit (see
the inset of Fig.~\ref{f.hist}(a)), give very strong support for the
existence of a discontinuous transition. Due to the large numerical
effort involved in calculating the local voltages~\cite{CompCost}, we
refrained from mapping in detail the full transition surface in the
three-dimensional $(p,q,\gamma)$ phase diagram. The important
conclusion we extract from Fig.~\ref{f.hist} is that also for
$\gamma\neq 0$ (in this case $\gamma=0.1$) one can find an explosive
metallic breakdown.

In summary, we have discovered that the metallic breakdown due to
pollution with metallic powder can become explosive if the inhibition
of adsorption due to a local electric field becomes too strong. In the
case $\gamma=0$, i.e., when the details of the field strength become
unimportant, the model can be solved exactly, representing one of the
rare examples where an explosive percolation transition at a finite
$p_c$ can be proven to exist. The physical reason why the metallic
breakdown becomes abrupt is the same as for other percolation
models~\cite{Achlioptas2009,Ziff09,DSouza10,daCosta10,Riordan11,Araujo2010,
Cho13,Chen11,Nagler11,Moreira2010,Cho2009,Reis2012,Araujo2011,Andrade11},
namely, the avoidance in occupying bonds that locally feel an electric
field suppresses the appearance of an infinite cluster. This effect
gets more pronounced the more large local fields matter, i.e., the
more cutting bonds are particularly punished.

The finding of the explosive metallic breakdown phenomenon may explain
the difficulties in predicting the failure of electronic circuits. It
might on the other hand also help mitigating the problem by working
under conditions corresponding to non-critical regions in phase space.
It is therefore interesting to further explore the present model and,
in particular, to incorporate into it more empirical information about
real IC circuits.

\begin{acknowledgments}
  We thank the Brazilian Agencies CNPq, CAPES, and FUNCAP, the
  National Institute of Science and Technology for Complex Systems in
  Brazil, the European Research Council (ERC) Advanced Grant
  319968-FlowCCS for financial support, and the Portuguese Foundation
  for Science and Technology (FCT) under Contracts IF/00255/2013,     
  PEst-OE/FIS/UI0618/2014, and EXCL/FIS-NAN/0083/2012.
\end{acknowledgments}

\end{document}